\documentclass[prb]{revtex4}

\usepackage{graphicx}

\usepackage{dcolumn}

\def\dps{\displaystyle}

\def\dps{\displaystyle}

\def\ot{\stackrel{\dps{\otimes}}{,}}

\def\ot{\stackrel{\dps{\otimes}}{,}}

\begin{document}

\title{Hamiltonian formalism of the DNLS equation with nonvanished boundary value}

\author{Hao Cai}
 \email{hcai@vip.sina.com}
\affiliation{%
Department of Physics, Wuhan University, Hubei 430072, People's Republic of China}%

\author{Nian-Ning Huang}
\affiliation{Department of Physics, Wuhan University, Hubei 430072, People's Republic of China}%

\date{\today}

\begin{abstract}
Hamiltonian formalism of the DNLS equation with nonvanishing boundary value is developed by the standard procedure.
\end{abstract}

\pacs{05.45.Yv, 42.65.Tg, 47.65.+a}

\maketitle

\section{Introduction}

From the general view point, the complete integrability of a nonlinear equation means it describes a multi-periodic system, that is, a Hamiltonian system with action-angle variables as canonical conjugate variables\cite{r1}. In the case of complex field equation, such as NLS equation, one can introduce field density and its canonical conjugate momentum density in usual sense. For real equation, such as KdV equation, to formulate Hamiltonian formalism it has been introduced an alternative form of Poisson bracket for real field densities at two points\cite{r2,r3}. Furthermore, the time-dependence of angle variables derived from its poisson bracket with action variables must be same with that derived from the inverse scattering transform, which has not been paid much attention to.

The derivative nonlinear Schr\"{o}dinger (DNLS) equation was proposed to describe nonlinear Alfv\'{e}n waves in plasma\cite{AR,MSR}. In the case of vanishing boundary, it was solved by the inverse scattering transform(IST)\cite{DJK-ACN}, or other approaches\cite{AN-HHC,NNH-ZYC,HS}. And the complete integrability of it was shown by Kundu in Ref.[10] through r-s matrix formalism. In the other case of non-vanishing boundary, the DNLS equation was discussed by some authors in terms of the usual spectral parameter\cite{TK-HI,TK-NK-HI}. The multi-value problem of square root appears, and then the Riemann surface has to be introduced, which leads to complexity and ambiguity in the derivation. As a result, the affin parameter is introduced to clarify the multi-value problem\cite{EM,EM-TH,XJC-WKL}, based upon which the Hamiltonian formalism of DNLS equation with non-vanishing boundary condition can be formulated naturally. We perform it in this work, and the effect of the linear coordinate transformation on the Hamiltonian theory is shown.

\section{poisson bracket}
The DNLS equation with non-vanishing boundary condition (DNLS$^+$ equation) is generally expressed as\cite{AR,MSR}
\begin{equation}\label{1-1}
iu_t-u_{xx}+i(|u|^2u)_x=0,
\end{equation}
where $u$ is complex and $|u|\to\rho$ in the limit of $x\to\pm\infty$. Now a particular form of Poisson bracket is proposed
\begin{equation}\label{1-2}
\{u(x),\overline{u(y)}\}=\frac{1}{2}\{\partial_x-\partial_y\}\delta(x-y).
\end{equation}
For two quantities $Q$, $R$, the Poisson bracket is
\begin{equation}\label{1-3}
\{Q,R\}=\int\int dxdy
\Big(\frac{\delta Q}{\delta u(x)}\frac{\delta R}{\delta
\overline{u(y)}}\{u(x),\overline{u(y)}\}
+\frac{\delta Q}{\delta \overline{u(x)}}\frac{\delta R}{\delta
u(y)}\{\overline{u(x)},u(y)\}\Big).
\end{equation}
Integrating by part, eq.(\ref{1-3}) becomes
\begin{eqnarray}\label{1-4}
\{Q,R\}=-\frac{1}{2}\int\int dxdy
\Big(\left\{\partial_x\frac{\delta Q}{\delta u(x)}\right\}\frac{\delta R}{\delta
\overline{u(y)}}
-\frac{\delta Q}{\delta u(x)}\left\{\partial_x\frac{\delta R}{\delta
\overline{u(y)}}\right\}\\
+\left\{\partial_x\frac{\delta Q}{\delta \overline{u(x)}}\right\}\frac{\delta R}{\delta
u(y)}
-\frac{\delta Q}{\delta \overline{u(x)}}\left\{\partial_x\frac{\delta R}{\delta
u(y)}\right\}\Big)\delta(x-y).\nonumber
\end{eqnarray}
Thus the Hamiltonian equation is obtained
\begin{equation}\label{1-5}
u_t(x)=\{H,u(x)\},~~
H=\int dx {\cal H}(x),~~{\cal H}(x)=\frac{1}{2}|u|^4-iu_x\bar{u}.
\end{equation}

\section{Lax pair}
The first one of Lax equations is
\begin{equation}\label{2-1}
\partial_x F(x,t,\lambda)=L(x,t,\lambda)F(x,t,\lambda),
\end{equation}
where $\lambda$ is spectral parameter. $L$ is a $2\times 2$ matrix
\begin{equation}\label{2-2}
L=-i\lambda^2\sigma_3+\lambda U,~~U=u\sigma_++\bar u\sigma_-,
\end{equation}
where
\begin{equation}\label{2-3}
\sigma_+=\left(\begin{array}{cc}
0 & 1 \\
0 & 0
\end{array}\right),~~
\sigma_-=\left(\begin{array}{cc}
0 & 0 \\
1 & 0
\end{array}\right),
\end{equation}
and $\zeta$ is an auxiliary parameter such that
\begin{equation}\label{2-4}
\lambda=\frac{1}{2}(\zeta+\rho^2\zeta^{-1}),~~
\kappa=\frac{1}{2}(\zeta-\rho^2\zeta^{-1}).
\end{equation}

As  the asymptotic free Jost solution is $E(x,\zeta)=(I+\rho\zeta\sigma_2)e^{-i\lambda\kappa x\sigma_3}$, we define the Jost solutions
\begin{eqnarray}\label{2-5}
\Psi(x,\zeta)=\big(\tilde{\psi}(x,\zeta), \ \psi(x,\zeta)\big)\to E(x,\zeta),~~
&\mbox{as} \ x\to\infty,\\
\Phi(x,\zeta)=\big(\phi(x,\zeta), \ \tilde{\phi}(x,\zeta)\big)\to Q^{-1}(\alpha)E(x,\zeta),~~
&\mbox{as} \ x\to-\infty,\nonumber
\end{eqnarray}
where $Q(\alpha)=e^{i\frac{1}{2}\alpha\sigma_3}$. Then the monodramy matrix $T(\lambda)$ is given by
\begin{equation}\label{2-6}
T(\zeta)=\Psi^{-1}(x,\zeta)\Phi(x,\zeta), \ \
T(\zeta)=\left(\begin{array}{rr}a(\zeta)&\tilde{b}(\zeta)\\
b(\zeta)&\tilde{a}(\zeta)\end{array}\right),
\end{equation}
From (\ref{2-5}), $a(\zeta)$ can be analytical continued into
the first and the third quadrants
and $\tilde a(\zeta)$ the second and the fourth quadrants.
The continuous spectrum is composed of real $\zeta^2$, that is composed of real
$\zeta$ and of imaginary $\zeta$.

Similar to NLS$^+$ equation, there are several reduction transformation properties
\begin{equation}\label{4-1}
\tilde\psi(x,\bar\zeta)=\sigma_1\psi(x,\zeta),~~
\tilde\phi(x,\bar\zeta)=\sigma_1\phi(x,\zeta),
\end{equation}
\begin{equation}\label{4-2}
\tilde a(\bar\zeta)=a(\zeta),~~
\tilde b(\bar\zeta)=b(\zeta).
\end{equation}
and, under the transformation $\zeta\to\rho\zeta^{-1}$,
\begin{equation}\label{4-3}
\tilde\psi(x,\rho^2\zeta^{-1})=i\rho^{-1}\zeta\psi(x,\zeta),~~
\tilde\phi(x,\rho^2\zeta^{-1})=-i\rho^{-1}\zeta\phi(x,\zeta),
\end{equation}
\begin{equation}\label{4-4}
\tilde a(\rho^2\zeta^{-1})=a(\zeta),~~
\tilde b(\rho^2\zeta^{-1})=-b(\zeta).
\end{equation}
in which the second ones of (\ref{4-2}) and (\ref{4-4}) require $\lambda\kappa$ is real.

And there are some reduction properties only of DNLS$^+$ equation, for example,
since
\begin{equation}\label{4-5}
L(-\zeta)=\sigma_3L(\zeta)\sigma_3,
\end{equation}
we have
\begin{equation}\label{4-6}
T(-\zeta)=\sigma_3T(\zeta)\sigma_3,
\end{equation}
that is
\begin{equation}\label{4-7}
a(-\zeta)=a(\zeta),~~
b(-\zeta)=-b(\zeta).
\end{equation}

Since $a(\lambda)$ is assumed to have $N$ simple poles in the 1st and 3rd quadrants,
$a(\lambda)$ can be expressed as\cite{r1}
\begin{equation}\label{2-8}
a(\lambda)=\prod_{n=1}^N\frac{\zeta-\zeta_n}{\zeta-\bar\zeta_n}\frac{\bar\zeta_n}{\zeta_n}
\exp\left\{\frac{\zeta}{i2\pi}\int_{\Gamma}\frac{\ln|a(\zeta')|^2}{(\zeta'-\zeta)\zeta'}\right\},
\end{equation}
where the integral contour $\Gamma=(0,+\infty)\bigcup(0,-\infty)\bigcup(+i\infty,0)\bigcup(-i\infty,0)$ along the real and imaginary axis.

\section{poisson bracket for monodramy matrix}

Since $\partial_x\det\Psi(x,\zeta)=0$ and $\partial_x\det\Phi(x,\zeta)=0$, we have
\begin{equation}\label{5-1}
\det\Psi(x,\zeta)=\det\Phi(x,\zeta)=\det E(x,\zeta)=1-\rho^2\zeta^{-2},
\end{equation}
and thus
\begin{equation}\label{5-2}
\det T(\zeta)=1,~~a(\zeta)\tilde a(\zeta)-b(\zeta)\tilde
b(\zeta)=1,
\end{equation}
\begin{equation}\label{5-3}
T^{-1}(\zeta)=\left(\begin{array}{cc}
\tilde a(\zeta)&-\tilde b(\zeta)\\
-b(\zeta)&a(\zeta)\end{array}
\right)
\end{equation}
and so on.

From the Lax equation eq.(\ref{2-1}), one obtains

Introducing the usual direct product $\otimes$, the Poisson bracket of the monodramy matrix is defined as
\begin{equation}\label{6-0}
    \{T(\zeta)\ot T^{-1}(\zeta')\}_{ij,kl}=\{T(\zeta)_{ij},T^{-1}(\zeta')_{kl}\}
\end{equation}
Substituting eqs.(\ref{5-7})$\sim$(\ref{5-20}),
the explicit expression of eq.(\ref{6-0}) is
\begin{equation}\label{6-1}
\{T(\zeta)\ot T^{-1}(\zeta')\}=-\frac{1}{2}\int dx
{\Psi}^{-1}(x,\zeta){\Phi}^{-1}(x,\zeta')R{\Phi}(x,\zeta){\Psi}(x,\zeta')
\end{equation}
where
\begin{eqnarray}\label{6-2}
R=-\left(i2\lambda^3\sigma_{+}-\lambda^2\bar{u}\sigma_3\right)\otimes
(\lambda'\sigma_{-})
-(\lambda\sigma_{+})\otimes
\left(i2\lambda'^3\sigma_{-}+\lambda'^2u\sigma_3\right)\\
+\left(i2\lambda^3\sigma_{-}+\lambda^2u\sigma_3\right)\otimes
(\lambda'\sigma_{+})
+(\lambda\sigma_{-})\otimes
\left(i2\lambda'^3\sigma_{+}-\lambda'^2\bar{u}\sigma_3\right).\nonumber
\end{eqnarray}
Eq.(\ref{6-2}) is expressed in matrix with row $\{i'l'\}$ and column $\{j'm'\}$
\begin{equation}\label{6-4}
-\left(\begin{array}{cccc}
0&-\lambda^2\lambda'u&\lambda\lambda'^2u&0\\
-\lambda^2\lambda'\bar{u}&0&i2\lambda^3\lambda'+i2\lambda\lambda'^3&-\lambda\lambda'^2u\\
\lambda\lambda'^2\bar{u}&-i2\lambda^3\lambda'-i2\lambda\lambda'^3&0&\lambda^2\lambda'u\\
0&-\lambda\lambda'^2\bar{u}&\lambda^2\lambda'\bar{u}&0
\end{array}\right).
\end{equation}

\section{another direct product for poisson bracket}
In the usual method to formulate Hamiltonian theory, one considers\cite{r1}
\begin{equation}\label{7-1}
\partial_x\Big(\{\Psi^{-1}(\zeta)\Psi(\zeta')\}\otimes'\{\Phi^{-1}(\zeta')\Phi(\zeta)\}\Big).
\end{equation}
where another direct product $\otimes'$ is introduced
\begin{equation}\label{7-3}
A_{im}B_{lj}=(A\otimes' B)_{il,jm}
\end{equation}
From the first Lax equation, eq.(\ref{7-1}) becomes
\begin{equation}\label{7-2}
\Psi^{-1}(\zeta)\{L(\zeta')-L(\zeta)\}\Psi(\zeta')\otimes'\Phi^{-1}(\zeta')\Phi(\zeta)
+\Psi^{-1}(\zeta)\Psi(\zeta')\otimes'\Phi^{-1}(\zeta')\{L(\zeta)-L(\zeta')\}\Phi(\zeta)
\end{equation}
that is
\begin{equation}\label{7-4}
{\Psi}^{-1}(x,\zeta){\Phi}^{-1}(x,\zeta')W_0{\Phi}(x,\zeta){\Psi}(x,\zeta')
\end{equation}
where
\begin{eqnarray}\label{7-5}
W_0&=&i(\lambda^2-\lambda'^2)\{\sigma_3\otimes'I-I\otimes'\sigma_3\}\\
&&-(\lambda-\lambda')u\{\sigma_{+}\otimes'I-I\otimes'\sigma_{+}\}
-(\lambda-\lambda')\bar{u}\{\sigma_{-}\otimes'I-I\otimes'\sigma_{-}\}.\nonumber
\end{eqnarray}
Eq.(\ref{7-5}) may be written in the matrix form explicitly
\begin{equation}\label{7-6}
\left(\begin{array}{cccc}
0&-(\lambda-\lambda')u&(\lambda-\lambda')u&0\\
(\lambda-\lambda')\bar{u}&0&i2(\lambda^2-\lambda'^2)&-(\lambda-\lambda')u\\
-(\lambda-\lambda')\bar{u}&-i2(\lambda^2-\lambda'^2)&0&(\lambda-\lambda')u\\
0&(\lambda-\lambda')\bar{u}&-(\lambda-\lambda')\bar{u}&0
\end{array}\right)
\end{equation}
It is obvious that eq.(\ref{7-6}) is not proportional to eq.(\ref{6-4}),
which means other expression is necessary to construct the Hamiltonian formalism.

Considering
\begin{equation}\label{7-7}
\partial_x\Big(\{\Psi^{-1}(\zeta)\sigma_3\Psi(\zeta')\}\otimes'\{\Phi^{-1}(\zeta')\sigma_3\Phi(\zeta)\}\Big),
\end{equation}
it is equal to
\begin{eqnarray}\label{7-8}
\Psi^{-1}(\zeta)\{\sigma_3L(\zeta')-L(\zeta)\sigma_3\}\Psi(\zeta')\otimes'\Phi^{-1}(\zeta')\sigma_3\Phi(\zeta)\\
+\Psi^{-1}(\zeta)\sigma_3\Psi(\zeta')\otimes'\Phi^{-1}(\zeta')\{\sigma_3L(\zeta)-L(\zeta')\sigma_3\}\Phi(\zeta).\nonumber
\end{eqnarray}
also from the first Lax equation.
Similar to (\ref{7-4}), (\ref{7-8}) may be written in the form
\begin{equation}\label{7-9}
{\Psi}^{-1}(x,\zeta){\Phi}^{-1}(x,\zeta')W_3{\Phi}(x,\zeta){\Psi}(x,\zeta')
\end{equation}
where
\begin{eqnarray}\label{7-10}
W_3&=&i(\lambda^2-\lambda'^2)\{I\otimes'\sigma_3-\sigma_3\otimes'I\}\\
&&+(\lambda+\lambda')u\{\sigma_{+}\otimes'\sigma_3+\sigma_3\otimes'\sigma_{+}\}
-(\lambda+\lambda')\bar{u}\{\sigma_{-}\otimes'\sigma_3+\sigma_3\otimes'\sigma_{-}\},\nonumber
\end{eqnarray}
explicitly
\begin{equation}\label{7-11}
\left(\begin{array}{cccc}
0&(\lambda+\lambda')u&(\lambda+\lambda')u&0\\
-(\lambda+\lambda')\bar{u}&0&-i2(\lambda^2-\lambda'^2)&-(\lambda+\lambda')u\\
-(\lambda+\lambda')\bar{u}&i2(\lambda^2-\lambda'^2)&0&-(\lambda+\lambda')u\\
0&(\lambda+\lambda')\bar{u}&(\lambda+\lambda')\bar{u}&0
\end{array}\right)
\end{equation}

Defining
\begin{equation}\label{7-11a}
    \Delta_{\alpha}\equiv\lim_{L\to\infty}\Psi^{-1}(x,\zeta)\sigma_{\alpha}\Psi(x,\zeta')\otimes'
        \Phi^{-1}(x,\zeta')\sigma_{\alpha}\Phi(x,\zeta)\big|_{x=-L}^{x=L}
\end{equation}
there should be
\begin{equation}\label{7-11b}
    \{T(\zeta)\ot T^{-1}(\zeta')\}=f_0\Delta_0+f_3\Delta_3
\end{equation}
where two constant coefficients $f_0$ and $f_3$ are introduced, that is,
$f_0(\ref{7-6})+f_3(\ref{7-11})=(\ref{6-4})$. Comparison between
the corresponding elements of matrices in two sides yields
\begin{equation}\label{7-12}
-(\lambda-\lambda')f_0+(\lambda+\lambda')f_3=\lambda^2\lambda',~
(\lambda-\lambda')f_0+(\lambda+\lambda')f_3=-\lambda\lambda'^2.
\end{equation}
It is found
\begin{equation}\label{7-13}
f_0=-\frac{1}{2}\lambda\lambda'\frac{\lambda+\lambda'}{\lambda-\lambda'}, \ \ \
f_3=\frac{1}{2}\lambda\lambda'\frac{\lambda-\lambda'}{\lambda+\lambda'}.
\end{equation}

\section{explicit expression of poisson bracket of monodramy matrix}

Finally the Poisson bracket $\{T(\lambda)\ot T^{-1}(\lambda')\}$, i.e.,
\begin{equation}\label{11-1}
\left(\begin{array}{cccc}
\{a,\tilde a'\}&-\{a,\tilde b'\}&\{\tilde b,\tilde a'\}&-\{\tilde b,\tilde b'\}\\
-\{a,b'\}&\{a,a'\}&-\{\tilde b,b'\}&\{\tilde b,a'\}\\
\{b,\tilde a'\}&-\{b,\tilde b'\}&\{\tilde a,\tilde a'\}&-\{\tilde a,\tilde b'\}\\
-\{b,b'\}&\{b,a'\}&-\{\tilde a,b'\}&\{\tilde a,a'\}\end{array}
\right)
\end{equation}
is equal the sum of (\ref{8-7}), (\ref{9-3}) and (\ref{9-6}), for example
\begin{equation}\label{11-3}
    \{a,b'\}=\frac{1}{2}\lambda\lambda'
\left\{\frac{\lambda\lambda'-\rho^2}{\kappa\kappa'}\frac{\lambda+\lambda'}{\lambda-\lambda'+i0}
+\frac{\lambda\lambda'+\rho^2}{\kappa\kappa'}\frac{\lambda-\lambda'}{\lambda+\lambda'+i0}
\right\}ab'
\end{equation}
and
\begin{equation}\label{11-4}
\{\tilde a,b'\}=-\frac{1}{2}\lambda\lambda'
\left\{\frac{\lambda\lambda'-\rho^2}{\kappa\kappa'}\frac{\lambda+\lambda'}{\lambda-\lambda'-i0}
+\frac{\lambda\lambda'+\rho^2}{\kappa\kappa'}\frac{\lambda-\lambda'}{\lambda+\lambda'-i0}
\right\}ab'
\end{equation}
As $\zeta$ is pure imaginary, i.e. $\zeta=i\eta$, noticing $\delta(i\eta)=\delta(\eta)$, we have also the same results with (\ref{11-3}) and (\ref{11-4}).

\section{Action-angle variables in continuous spectrum}

From the inverse scattering transform, $a(\zeta)$ and $\tilde a(\zeta)$ are independent of t,
and $b(\zeta)$ and $\tilde b(\zeta)$ depend on $t$. And noticing the reduction transformation
properties (\ref{4-4}), and (\ref{4-7}), we may restrict ourselves to consider $\zeta$, $\zeta'>\rho$.
Thus from (\ref{11-3}) and (\ref{11-4}), there is
\begin{equation}\label{12-1}
    \{|a(\zeta)|^2,b(\zeta')\}=-i2\lambda^3\pi|a(\zeta)|^2b(\zeta)(1-\rho^2\zeta^{-2})^{-1}
    \delta(\zeta-\zeta')
\end{equation}
Angle variable $Q(\zeta)$ and action variable $P(\zeta)$ are chosen, respectively, to be
\begin{equation}\label{12-2}
    Q(\zeta)=\mbox{arg}b(\zeta)=\frac{1}{2i}\ln\frac{b(\zeta)}{\tilde b(\zeta)},~
    P(\zeta)=F(|a(\zeta)|^2)
\end{equation}
such that
\begin{equation}\label{12-3}
    \{P(\zeta),Q(\zeta')\}=-\delta(\zeta-\zeta')
\end{equation}
where the unknown function $F$ is to be determined by (\ref{12-1}).

Substituting (\ref{12-1}) into (\ref{12-3}), it is easy to find
\begin{equation}\label{12-4}
    F'(|a(\zeta)|^2)2\lambda^3\pi|a(\zeta)|^2(1-\rho^2\zeta^{-2})^{-1}=1
\end{equation}
Thus the action variable $P(\zeta)$ is chosen as
\begin{equation}\label{12-5}
    P(\zeta)=F(|a(\zeta)|^2)=\frac{1-\rho^2\zeta^{-2}}{2\lambda^3\pi}\ln|a(\zeta)|^2
\end{equation}

\section{Action-angle variable in discrete spectrum}

From inverse scattering transform we know that $\lambda_n$, zero of $a(\lambda)$, is independent of time,
and $b_n$ is dependent on time periodically. Hence we need Poisson bracket of $\zeta_m$ with $b_n$,
and of them with $a(\zeta)$ and $b(\zeta)$. From (\ref{2-8}) and (\ref{12-1}) we obtain
\begin{equation}\label{13-1}
\{{\rm ln}\check{a}(\zeta),b({\zeta'})\}+\sum_m\left(
\frac{\{\bar{\zeta}_m,b({\zeta'})\}}{\zeta-\bar{\zeta}_m}
-\frac{\{\zeta_m,b({\zeta'})\}}{\zeta-\zeta_m}\right)
=b({\zeta'})\left\{\lambda\lambda'\frac{\lambda+\lambda'}{2}\frac{\lambda\lambda'-\rho^2}{\kappa\kappa'}
\frac{1}{\lambda-\lambda'+i0}+\lambda\lambda'\frac{\lambda-\lambda'}{2}\frac{\lambda\lambda'+\rho^2}{\kappa\kappa'}
\frac{1}{\lambda+\lambda'+i0}\right\}
\end{equation}
If $\zeta=\zeta_m$, then $\lambda_m={\rm Re}\zeta_m$ is real,
$\lambda_m-\lambda'+i0\neq 0$ and $\lambda_m+\lambda'+i0\neq 0$
since $\lambda'$ is real. The right hand side indicates
$\lambda_m$ is not a pole, that is, $\{\zeta_m, b({\zeta'})\}=0$.
Similarly, we have $\{\bar{\zeta}_m, b({\zeta'})\}=0$. Then by the
standard procedure, similar to (\ref{13-1}), we have
\begin{equation}\label{13-2}
\{{\rm ln}\check{a}(\zeta),b_n\}+\sum_m\left(
\frac{\{\bar{\zeta}_m,b_n\}}{\zeta-\bar{\zeta}_m}
-\frac{\{\zeta_m,b_n\}}{\zeta-\zeta_m}\right)
=b_n\left\{\lambda\lambda_n
\frac{\lambda+\lambda_n}{2}\frac{\lambda\lambda_n-\rho^2}{\kappa\kappa_n}
\frac{1}{\lambda-\lambda_n}+\lambda\lambda_n
\frac{\lambda-\lambda_n}{2}\frac{\lambda\lambda_n+\rho^2}{\kappa\kappa_n}
\frac{1}{\lambda+\lambda_n}\right\}
\end{equation}
The right hand side has a pole at $\zeta=\zeta_n$, noting that
$\lambda-\lambda_n=\frac{1}{2}(\zeta-\zeta_n)(1-\rho^2\zeta^{-1}\zeta_n^{-1})$
we obtain
\begin{equation}\label{13-3}
    \{\zeta_m,b_n\}=-2\lambda_n^3(1-\rho^2\zeta_n^2)^{-1}b_n\delta_{mn}
\end{equation}
This result is similar to that of the Hamiltonian theory for other nonlinear equations,
but there is something different because $\bar\zeta_n\neq \rho^2\zeta_n^{-1}$,
that is, $\zeta-\bar\zeta_n$ is not a factor of $\lambda-\lambda_n$, which means
\begin{equation}\label{13-3a}
    \{\bar{\zeta}_m,b_n\}=0
\end{equation}
Furthermore, we have also
$\{a(\zeta), \zeta_m\}=0$ and $\{\zeta_n,\zeta_m\}=0$.

In the discrete spectrum case, the angle variable is
\begin{equation}\label{13-4}
Q_n={\rm ln}b_n
\end{equation}
and the action variable is assumed to be $P_m=G(\zeta_m)$,
where $G$ is an unknown function.
The Poisson bracket of them must be
$\{P_m,Q_n\}=-\delta_{mn}$,
and then, noticing (\ref{13-3}) and (\ref{13-4}), we have
\begin{equation}\label{13-5}
G'(\zeta_m)2\lambda_m^3(1-\rho^2\zeta_m^{-2})^{-1}=1
\end{equation}
We thus obtain
\begin{equation}\label{13-6}
P_m=G(\zeta_m)=-\frac{1}{2\lambda_m^2}
\end{equation}

\section{Conservative quantities}

Since the first one of the Lax pair of DNLS$^+$ equation is the
same as that of NLS$^+$ equation, the conservative quantities are
the same. We have
\begin{equation}\label{14-1}
{\rm ln}a(\zeta)=\sum_n{\rm
ln}\left(\frac{\zeta-\zeta_n}{\zeta-\bar{\zeta}_n}\frac{\bar\zeta_n}{\zeta_n}\right)
-\frac{\zeta}{i2\pi}\int_{\Gamma} d{\zeta'}\frac{{\rm
ln}|a({\zeta'})|^2}{({\zeta'}-\zeta)\zeta'}
\end{equation}
Since $a(\zeta)$ is a constant in time, all terms in expansion of
$|\zeta|\to\infty$ are constant, for example,
\begin{equation}\label{14-2}
    I_0=\sum_n2\ln\frac{\bar\zeta_n}{\zeta_n}
+\frac{1}{i\pi}\int_{\Gamma_+}
d{\zeta'}\frac{1}{\zeta'}{\rm ln}|a({\zeta'})|^2
\end{equation}
\begin{equation}\label{14-4}
    I_2=\sum_m(\bar{\zeta}_m^2-\zeta_m^2)
+\frac{1}{i\pi}\int_{\Gamma_+} d{\zeta'}\zeta'{\rm
ln}|a({\zeta'})|^2
\end{equation}
etc., where we have taken account of $|a(\zeta)|^2=|a(-\zeta)|^2$
and the condition that $-\zeta_m$ is a zero of $a(\zeta)$ as
long as $\zeta_m$ is a zero of $a(\zeta)$, see (\ref{2-8}). The Hamiltonian is assumed to be
\begin{equation}\label{14-6}
H=iI_2-i\eta I_0
=\sum_mi[(\bar{\zeta}_m^2-\zeta_m^2)-2\eta(\ln\bar{\zeta}_m-\ln\zeta_n)]
+\frac{1}{\pi}\int_{\Gamma_+} d{\zeta'}(\zeta'-\eta\frac{1}{\zeta'}){\rm
ln}|a({\zeta'})|^2
\end{equation}
where the contour $\Gamma_+$ is along the 1st quadrant, and $\eta$ is a real constant we shall determine. The integral
domain $(0,\rho)$ can be transformed to $(\rho, \infty)$ by
${\zeta'}\to\rho^2{\zeta'}^{-1}$, the integral part is now given
by
\begin{equation}\label{14-7}
H_{int}=\frac{1}{i\pi}\int_{\Gamma_+\cup \{|\zeta'|>\rho\}} d{\zeta'}
\left\{(\zeta'+\rho^4{\zeta'}^{-3})
-2\eta\zeta'^{-1}\right\}{\rm ln}|a({\zeta'})|^2
\end{equation}
From (\ref{12-1}) the $H_{int}$ must involve a factor
$(1-\rho^2{\zeta'}^{-2})$, it is easily seen that if and only if
$\eta=\rho^2$ the factor in bracket can be factored as
\begin{equation}\label{14-8}
(\zeta'+\rho^4{\zeta'}^{-3})-2\eta\zeta'^{-1}
=(\zeta'-\rho^2{\zeta'}^{-1})(1-\rho^2{\zeta'}^{-2})
\end{equation}
In this choice we obtain from (\ref{12-1})
\begin{equation}\label{14-9}
\{H_{int},b(\zeta)\}=-i4\lambda^3\kappa b(\zeta)
\end{equation}
From eqs.(\ref{13-3}) and (\ref{13-3a}), we have
\begin{equation}\label{14-10}
\{H_{dis},b_n\}=-i4\lambda_n^3\kappa_n b_n
\end{equation}
where $H_{dis}$ is the summation part. From eqs.(\ref{14-9}) and (\ref{14-10}), there are
\begin{equation}\label{15-1}
    b(t,\zeta)=b(0,\zeta)e^{i4\lambda^3\kappa t},~b_n(t)=b_n(0)e^{i4\lambda_n^3\kappa_nt}
\end{equation}
In such a choice, the second one of the Lax pair should be
\begin{equation}\label{15-2}
    M=-i2\lambda^4\sigma_3+2\lambda^3U-i\lambda^2(U^2-\rho^2)\sigma_3+\lambda U(U^2-\rho^2)-i\lambda U_x\sigma_3
\end{equation}
which is different from the usual form.
As a result, the compatibility condition gives
\begin{equation}\label{15-3}
    iu_t-u_{xx}+i[(|u|^2-\rho^2)u]_x=0
\end{equation}
which differs from the usual form of the DNLS$^+$ equation (\ref{1-1}) by a Galileo transformation
\begin{equation}\label{15-4}
    t'=t,~x'=x-\rho^2t
\end{equation}
Correspondingly, different from eq.(\ref{1-5}), here the Hamiltonian density is chosen as
\begin{equation}\label{15-5}
    {\cal H}(x)=\frac{1}{2}(|u|^2-\rho^2)^2-iu_x\bar{u}
\end{equation}

Thus we formulate the complete Hamiltonian theory for the DNLS equation with non-vanishing boundary conditions. At the end, the linear Galileo transformation is introduced to take the time dependence of Angle variables derived from its Poisson bracket with the Hamiltonian compatible with that derived from the second Lax equation. Such a transformation is also beneficial to the procedure of solving the DNLS equation, which should be shown in the following work.


\begin{thebibliography}{100}
\bibitem{r1} L. D. Faddeev and L. A. Takhtajan, Hamiltonian Methods in the Theory of Solitons (Springer, Berlin, 1987).
\bibitem{r2} V.E.Zakharov and A.B.Sabat, Sov. Phys. JETP, {\it 34}, 62(1972)
\bibitem{r3} V.E.Zakharov and L.D.Faddeev, Funkz. Anal. Priloz., {\bf 5}, No.4,
18 (1971, Russian)
\bibitem{AR} A. Rogister, Phys. Fluids 14, 2733 (1971).
\bibitem{MSR} M. S. Ruderman, J. Plasma Phys. 67, 271 (2002).
\bibitem{DJK-ACN} D. J. Kaup and A. C. Newell, J. Math. Phys. 19, 798 (1978).
\bibitem{AN-HHC} A. Nakamura and H. H. Chen, J. Phys. Soc. Jpn. 49, 813
(1980).
\bibitem{NNH-ZYC} N. N. Huang and Z. Y. Chen, J. Phys. A 23, 439 (1990).
\bibitem{HS} H. Steudel, J. Phys. A 36, 1931 (2003).
\bibitem{Kundu} A. Kundu, J. Phys. A 21, 945 (1988)
\bibitem{TK-HI} T. Kawata and H. Inoue, J. Phys. Soc. Jpn. 44, 1968 (1978).
\bibitem{TK-NK-HI} T. Kawata, N. Kobayashi, and H. Inoue, J. Phys. Soc. Jpn. 46, 1008 (1979).
\bibitem{EM} E. Mj\O lhus, Phys. Scr. 40, 227 (1989).
\bibitem{EM-TH} E. Mj\O lhus and T. Hada, in Nonlinear Waves and Chaos in Space Plasmas, edited by T. Hada and H. Matsumoto (Terrapub, Tokyo, 1997), p. 121.
\bibitem{XJC-WKL} X. J. Chen and Wa Kun Lam, Phys. Rev. E, 69, 066604 (2004)
\end{thebibliography}
\end{document}